\documentclass[doublecol,comment]{epl2} 
\newcommand{\rem}[1]{}
\title{Comment on ``The role of wetting heterogeneities in the meandering instability of a partial wetting rivulet''}
\shorttitle{Comment on ``... the meandering instability of a partial wetting rivulet''} %Insert here a short version of the title if it exceeds 70 characters

\author{N. Fathi\inst{1} \and K. Mertens\inst{2} \and V. Putkaradze\inst{3} \and P. Vorobieff\inst{1}}
\shortauthor{N. Fathi \etal}

\institute{                    
  \inst{1} The University of New Mexico - Albuquerque, NM 87131, USA\\
  \inst{2} LeapMotion, Inc. - 333 Bryant St. Ste. LL150, San Francisco, CA 94017, USA\\
  \inst{3} The University of Alberta - Edmonton, AB T6G2G1, Canada
}

\begin{document}

\maketitle

Rivulets \cite{rivulet1} and their meandering on a partially wetting surface \cite{culkin}
present an interesting problem, as complex behavior arises from a deceptively
simple setup.
Recently Couvreur and Daerr \cite{daerr-new} suggested that meandering 
is caused by an instability developing as the flow rate $Q$ increases to a critical value $Q_c$, 
with stationary (pinned) meandering being the final state of the flow.
% They assert that these claims are supported 
% by both theory 
% and experiment. 
We tried to verify this assertion experimentally, but instead produced results contradicting the claim of
Ref.~\cite{daerr-new}.
The likely reason behind the discrepancy is the persistence of flow-rate perturbations.
Moreover, the theory presented in this paper cannot reproduce
the states as considered and disagrees with other theories \cite{homsy,jfm08,daerr-old}.

First, we tried reproducing the critical flow rate precipitating meandering as reported 
\cite{daerr-new}. We were unable to do so with two carefully constructed 
experimental arrangements (one at the University of New Mexico, another at the University of North Carolina), 
both using the same substrate (glass), 
same fluid (water), and same flow parameters as the experiments of Ref.~\cite{daerr-new}, with the fluid supply 
following the design described in our previous work \cite{jfm08}.
The stationary pattern that emerged was a non-meandering, straight flow
over the span exceeding 2~m. This applies to flow rates 0.2--8 ml/s, while the range of flow rate of 
Ref.~\cite{daerr-new} was 0.2--1.8 ml/s. 
The likely cause of this difference is the  
``constant level tank'' Couvreur and Daerr employ: a constant (on average) level of fluid in the tank by itself does not 
guarantee that the \emph{instantaneous} flow rate is constant (only the average), and the flow meandering is keenly sensitive to even modest
flow rate $Q$ perturbations, as discussed in \cite{jfm05,jfm08,prl08}. 

In a tank with a source of velocity fluctuations near the bottom (e.g., a pump), these fluctuations 
rapidly decay away from the source (consider exponential decay in Stokes' second problem). Thus the top (far) boundary 
well may appear unperturbed, while the discharge rate from the bottom of the tank is affected. 

%It 
%is ironic that Couvreur and Daerr write: ``The fact that several of the preceding studies [9,14] report 
%meandering despite using the same constant level tank to minimize flow rate fluctuations, is not explained.''
%Ref.~\cite{prl08} states: ``It likely follows that universal claims
%in the literature about the inevitability of meandering were
%caused by the nearly unavoidable presence of flow-rate
%fluctuations in most flows.''  However, Refs.~\cite{jfm08,prl08,jfm05} all describe the 
%apparatus
%necessary to get rid of $Q$ fluctuations. 
% The space here does not allow for extensive quotations,
% but we will gladly point the authors to the relevant paragraphs, 
% should they continue to have trouble finding them.  

%To further examine the effects of the flow rate on the meandering/non-meandering character of the flow, 
%we conducted several experiments with steady and transient flow rates. For the flow rate of 5.7 ml/sec (Fig.~\ref{fig.1}), %the flow is perfectly straight, whereas Fig.~3 from Ref.~\cite{daerr-new} lists a range of 
%``critical meandering rates'' varying between 0.2 and 1.8 ml/s for transition to meandering. 

\begin{figure}
\centerline{\includegraphics{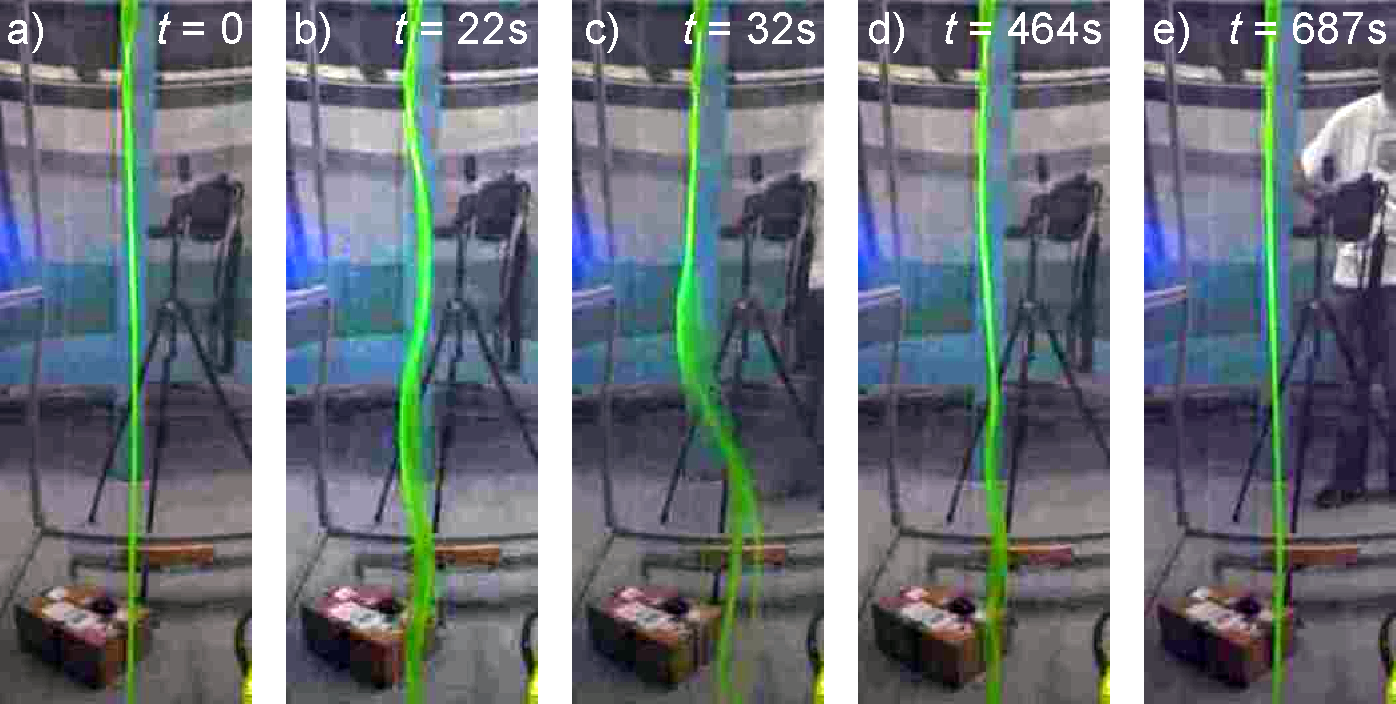}} 
\caption{a) - Straight rivulet at $Q=5.7$~ml/s (exists above meandering thresholds $Q_c$ 
predicted in Ref.~\cite{daerr-new}, requires no surface ``preparing'' to develop). 
 b) -- e) - image sequence showing destruction and re-emergence of the straight 
rivulet following a sequence of three 1~s flow-rate pulses with 1.5~s intervals (first pulse at $t=0$). 
Vertical image extent is 2~m.}
\label{fig.1}
\end{figure}

Any $Q$ variation (\emph{e.g.}, $Q$ increase) \emph{can} temporarily destabilize a rivulet
and mislead an observer into believing it has precipitated meandering. We have recorded \cite{aps12} 
transient meandering in response to $Q$ increase or decrease between constant levels (how slow a rate
change should be not to trigger meandering would be an interesting subject for further study).  
A stationary flow can be driven to 
meander with a short sequence of rate fluctuations (Fig.~\ref{fig.1}) retaining average $Q$ and tank fluid level. In all these cases, we see almost immediate transition to meandering. However, after $Q$ becomes constant, the straight flow typically resumes, often in the matter of minutes, although sometimes it takes longer.
%Thus, if the flow perturbations are allowed to occur more often 
%than once every 4 minutes, the rivulet will keep meandering continuously. 
Moreover, a ``pinned'' 
meandering pattern can be destroyed by a $Q$ increase, once again followed by formation of a straight rivulet \cite{aps12}. 
%The time necessary to record the experiments must be substantially longer than 5 minutes that the authors used to conclude %the convergence to a steady state. 

The theory presented in Ref.~\cite{daerr-new} may also contain unclear expositions, errors, or inconsistencies. 
It would help to specify that the axis of the independent coordinate $X$ ($x$ in dimensionless form) is normal to the direction of the rivulet, rather than pointing downstream.  Setting $X$ to be the downstream direction and $h$ to be the deviation from the centerline (a common notation in the field) leads to apparent inconsistencies. The most important of these are as follows. First, in the balance of forces, terms that are normal to each other would be equated. Second, Eq.~3 for $h'''(X)$ would contain stream curvature on the right hand side, which cannot be taken as constant (as it is taken in \cite{daerr-new}). Third, as a simple calculation shows, there would be no consistent limit for $h(X)$ far downstream, as all solutions to Eq.~3 would develop a singularity corresponding to the stream flowing sideways. Thus we will regard the $X$-direction as transversal to the rivulet.  Even then, questions still persist about the validity  of the key 
element of the theory,  namely the discussion following Eq~3. 
 
Indeed, Eq.~3  for the cross-sectional profile $h(x)$ is  $h'''(x) = - \alpha h^4(x) $, with $\alpha$ being a constant depending on the parameters of the flow. In Eq.~4, a cubic, area-preserving, polynomial expansion to this equation is assumed, stating $h = \theta_s/2 (1-x^2)(1+Ax/3)$, where $\theta_s$  is the (tangent of) contact angle of the equilibrium profile and $A$ is the asymmetry parameter. This expansion, applied for 
$-1<x<1$, does not represent a true solution of the differential equation and introduces artificial constraints, such as an 
additional linear dependence between derivatives evaluated at $x = \pm 1$.  
% We would like to point out that such ansatz describes an area-preserving solution of any equation $h'''(x)=F(h,x)$ independent of the right hand side $F$ for small $x$ and does not satisfy Eq.~3. 
%Implicit assumptions follow this ansatz that are not rooted in physics, such as linear connection between derivatives at the edges. 
However, the most serious criticism concerns the assumption of constancy of the cross-section,  used as an additional condition to determine the form of polynomial (p.~2 of the paper).  The constancy of area, \emph{i.e.} $\int_{-1}^1 h(x) \mbox{d} x$, does not follow from the equation for $h(x)$ or from any other physical principle.  The quantity that is constant for all steady states is the fluid flux  through a given cross-section and \emph{not} the area.   Since the theory is based on the \emph{deviations} from the straight rivulet, inaccurate description of variations of cross-sectional profile casts doubt on the validity of the whole theory.  

{In addition, the polynomial ansatz itself introduces errors, as shown in fig. 2, providing an exact solution of eq. (3) (solid line) and the corresponding polynomial (dashed line) satisfying exactly the same boundary conditions for 
$h(-1)$ and $h'(-1)$ with $h''(-1)$ for eq. (3) chosen so $h(1)=0$, for a particular value of $\theta_s$. Clearly, the area under the solid curve is  greater than that under the dashed curve. The difference between the areas changes with the 
{contact angles, tending to increase when one of the contact angles is increasing, precisely where the theory is applied}. This area  
mismatch is also present if one were to assign the same derivatives to the exact solution and polynomial ansatz at both ends. }
\begin{figure}
\centerline{\includegraphics[width=2.0in]{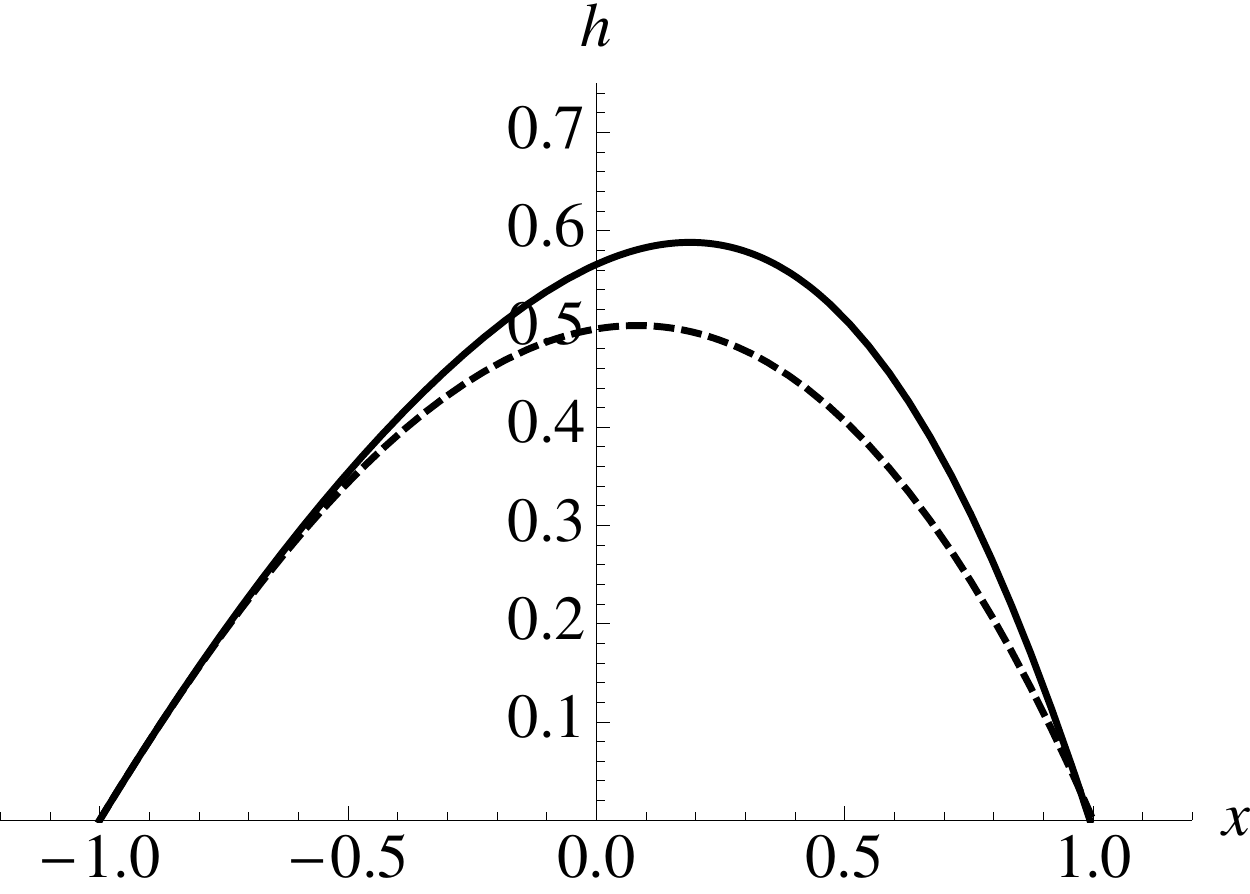}} 
\caption{Numerical solution for $h$ using the full equation (solid line) and polynomial ansatz (dashed line).}
\label{fig:wrong_eq}
\end{figure}  
%\color{black} 

Some of the difference in the interpretation of results may come from a 
different approach to time scales. The time necessary for straight
rivulet to establish in our experiment varies from minutes to hours. 
During the transition to straight steady state (which is sustainable indefinitely -- for days), 
meandering patterns that appear may look stationary, but destabilize in a matter of minutes.

Finally, let us comment on the interpretation of theoretical results after Eq.~8. The authors take the 
RMS of curvature and use it as a length parameter in the problem. Our previous measurements show that for meandering, all measurable quantities, properly averaged, satisfy a power law distribution \cite{prl08}. 
In the light of this observation, any curvature RMS is likely to depend on the particular cutoff and numerical procedure and thus may not be suited for use as a robust length scale. 

%We are quite saddened by the fallacies in many statements 
%of Ref.~\cite{daerr-new}, lack of rigour or
%attention to detail, both in experimental and theoretical 
%part, as well as a rather superficial review of previous %literature. 
%We hope that in the future, our colleagues  will report and analyze 
%their results in a more diligent and thorough %way. 

The problem of rivulet meandering in a variety of settings is interesting both as an example of 
a simple flow with surprisingly complex behavior, and because of its practical importance in a
wide variety of areas. However, this very complexity requires that the problem is treated rigorously and
with attention to detail -- both in theoretical consideration and in experimental approach.

\end{document}